\documentclass[aps,prl,twocolumn,groupedaddress,showpacs,floatfix]{revtex4}
\usepackage{graphicx}

\begin{document}

\title{Pump-Probe Experiments on the Single-Molecule Magnet $\rm Fe_8$: Measurement of Excited Level Lifetimes}

\author{S. Bahr$^1$, K. Petukhov$^{1}$\footnote{presently at Physikalisches Institut III, Universit$\ddot{\rm a}$t Erlangen, Germany}, V.Mosser$^2$, W. Wernsdorfer$^1$}

\affiliation{
$^1$Institut N\'eel, associ\'e \`a l'UJF, CNRS, BP 166, 38042 Grenoble Cedex 9, France\\
$^2$Itron France, 76 avenue Pierre Brossolette, 92240 Malakoff , France}

\date{submitted 1 March 2007}

\begin{abstract}
We present magnetization measurements on the single molecule magnet $\rm Fe_8$ in the presence of pulsed microwave radiation. 
A pump-probe technique is used with two microwave pulses with frequencies of 107~GHz and 118~GHz and pulse lengths of several nanoseconds to study the spin dynamics via time-resolved magnetization measurements using a Hall probe magnetometer. 
We find evidence for short spin-phonon relaxation times of the order of 1~$\rm \mu$s.
The temperature dependence of the spin-phonon relaxation time in our experiments is in good agreement with previously published theoretical results.
We also established the presence of very short energy diffusion times, that act on a timescale of about 70~ns.
\end{abstract}

\pacs{75.50.Xx, 75.60.Jk, 75.75.+a, 76.30.-v}

\maketitle

Single molecular magnets (SMMs) are the novel class of materials, where identical iso-oriented magnetic molecules are  regularly assembled in large crystals. Each of the molecules is built of superexchange-coupled magnetic metal ions; at low temperatures the coupling is so strong that the whole molecule can form a ground state described by a single net spin $S$.\cite{novak:1995,barra:1996,caciuffo:1998}
By applying a magnetic field, the net (giant) spin of the molecule can be reversed; such uniaxial magnetic bistability is evidenced by hysteretic magnetization measurements. Quantum tunnelling of magnetization (QTM) though the magnetic anisotropy barrier $E=DS^2$, where $D$ is the uniaxial anisotropy parameter, is established by the presence of steps in hysteresis loops of SMMs at millikelvin temperatures.\cite{friedman:1996,thomas:1996,Sangregorio:1997,wernsdorfer:science1999,sorace:2003}
At higher temperatures (few kelvin), the thermally activated relaxation can drive the molecule from the
ground state spin orientation $S$ over the barrier $E$ to the opposite orientation $-S$.\cite{novak:1995,barra:1996,caciuffo:1998,friedman:1996,thomas:1996,Sangregorio:1997,wernsdorfer:science1999,sorace:2003}
Due to their unique quantum properties and magnetic bistability, SMMs are currently considered as promising candidates for a variety of exciting applications, such as high-density magnetic data storage, quantum computation and magnetoelectronics.\cite{leuenberger:2001,troiani:2005,ardavan:2007,romeike:2007}
In order to develop these applications, it is important to be able to control the spin dynamics. 
For that purpose, we need to understand how the SMMs interact with their environment.  
In particular, it is essential to know the lifetimes of the exited spin states $\tau_m$, that define the spin-phonon relaxation time $T_1$.  
The use of microwave radiation, that induces selective transitions from the ground state to the excited states, can provide a direct access to the spin dynamics. 
Such experiments by means of continuous-wave electron spin resonance (ESR) \cite{barra:1996,mukhin:2001} and pulsed microwaves  time-resolved magnetometry \cite{wernsdorfer:epl2004,bal:2004,barco:2004,petukhov:2005,cage:MMM2005,petukhov:2007} were performed during last years in order to get a deeper insight to the spin relaxation time $T_1$ and decoherence time $T_2$. 
In these experiments it was impossible to measure $T_1$ directly because of the very strong phonon-bottleneck effect in $\rm Fe_8$, that screens out the fast relaxation processes.\cite{petukhov:2007}

\begin{figure}[t]
\includegraphics[height=3.2in]{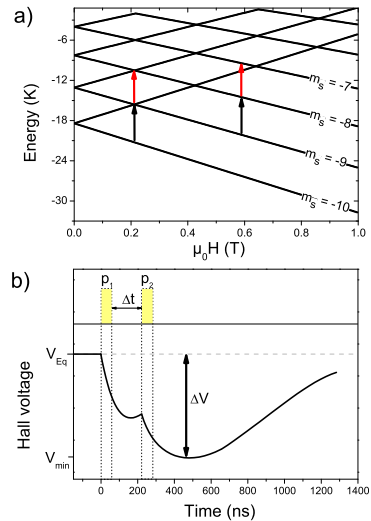}
\caption{\label{fig1} (a) Zeeman diagram of the molecular magnet $\rm Fe_8$. 
For the lowest spin states the transitions for the microwave frequencies $\rm f=118$~GHz and $\rm f=107$~GHz occure at a magnetic field of $\rm \mu_0 H\approx 0.2$~T.
(b) Schematic view of a typical pump-probe experiment at T=2~K. 
Two microwave pulses $\rm p_1$ and $\rm p_2$ separated by a delay $\Delta t$ excite the spin system and the magnetization of the sample, i.e. the Hall voltage decreases, reaches a minimum and relaxes back to the equilibrium value.
The amplitude $\Delta V = V_{eq}-V_{min}$ depends on the amount of spins excited by the two microwave pulses.
}
\end{figure}

A possibility to overcome this problem and to get insight in fast relaxation processes and spin dynamics on a nanosecond scale is the use of a time-resolved pump-probe technique.
A typical pump-probe experiment employs short pulses of two different microwave frequencies, that match the resonance condition between two energy levels at the same magnetic field value, as shown in Fig. 1a. 
The time resolution of the experiment is achieved by precise control of the delay time between the pump pulse and the probe pulse.
In this letter we use different two-pulse-techniques with microwave radiation in order to study the level lifetimes of excited spin states in the molecular magnet $\rm Fe_8$ as well as fast energy diffusion processes.

\begin{figure}[t]
\includegraphics[height=2.1in]{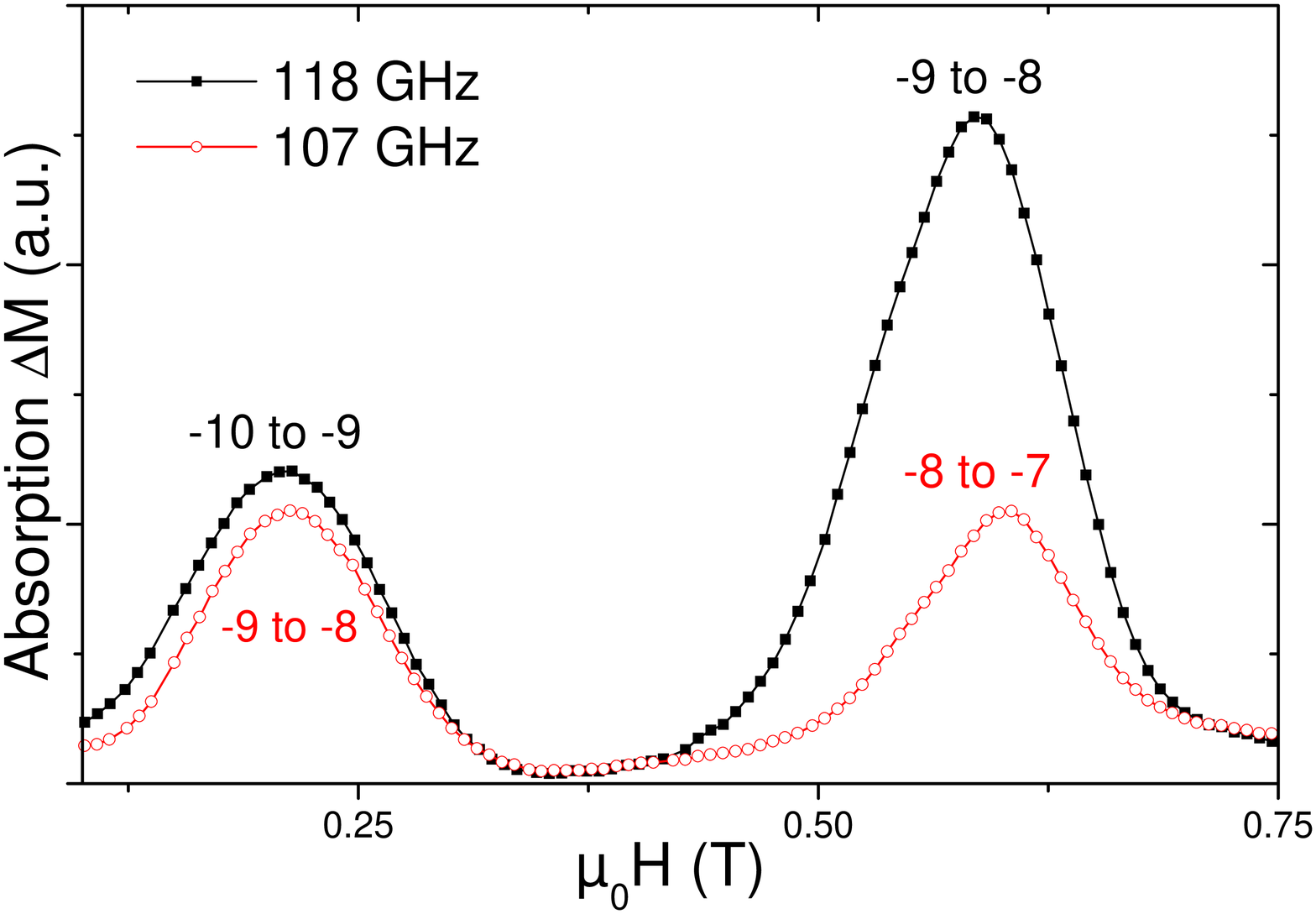}
\caption{\label{fig2}(Color online) Continuous-wave microwave absorption vs magnetic field measured via the decrease of the sample's magnetization at T=2~K.  According to the microwave frequency various transitions between spin states can be precisely observed, even with nanosecond microwave pulses. For microwave frequencies 118~GHz and 107~GHz the resonances appear at the same magnetic fields (c.f. Fig. \ref{fig1}a).
Note that only every fifth datapoint is shown.
}
\end{figure}

In our work we focus on the transitions between the ground state and the first excited states that occur for microwave frequencies of $f=118$~GHz and $f=107$~GHz at a magnetic field $\rm \mu_0 H = 0.2$~T, as can be seen in Fig. \ref{fig1}a.
When applying a short pulse of microwaves, the magnetization of the sample decreases during resonant excitation and after the end of the pulse it relaxes back to the equilibrium value (Fig. \ref{fig1}b). 
The microwave pulses have a length of typically 20 to 50~ns.
Therefore we can be sure that the sample's temperature is almost constant during such short excitation, as it was shown previously for much longer pulses.\cite{petukhov:2005}
The first microwave pulse excites spins from an energy level $m$ to an excited energy level $m+1$.
The second microwave pulse excites the spins from the excited level $m+1$ to an even higher level $m+2$ with a time delay $\Delta t$ in respect to the first microwave pulse.
The magnetization decrease $\Delta V$, due to the second pulse, only depends on the lifetime $\tau_{m+1}$ of the energy level $m+1$
For $\Delta t \ll \tau_{m+1}$ most of the spins pumped from $m$ to $m+1$ are still in the state $m+1$ and can be excited with the second pulse to the higher spin state $m+2$.
For $\Delta t \gg \tau_{m+1}$ most of the spins from the state $m+1$ already relaxed to the ground state $m$ and therefore the second pulse is not very efficient in exciting spins from  $m+1$ to $m+2$.
Thus the magnetization decrease $\Delta V$ as function of the delay time $\Delta t$ between the pulses provides information about the fast relaxation processes on a nanosecond scale.


Continuous microwaves were produced by mechanically tuneable Gunn oscillators with nominal output power of several milliwatts.
A PIN switch allowed us to produce very short pulses of microwaves down to 3~ns.
We chose a repetition rate of the pulse sequences of 5000 Hz to insure that the crystal stays at the cryostat temperature.
The microwaves propageated in an oversized circular waveguide and were focused with a conical end piece to the sample. The sample was placed into a $\rm ^4 He$ cryostat with its easy axis parallel to the external applied magnetic field.
The time-resolved magnetization measurements on the micrometer-sized sample were done with a Hall bar magnetometer.
The Hall bar with a 10 $\rm \mu$m arm width were fabricated from delta-doped AlGaAs/InGaAs/GaAs pseudomorphic
heterostructures grown by \emph{Picogiga} using molecular beam epitaxy.\cite{petukhov:2007}
The Hall voltage, that is of the order of magnitude of several $\rm \mu$V, is amplified by a low noise preamplifier with large frequency bandwidth $f_{BW}>100$~MHz.
The time-resolved magnetization measurements were averaged typically over 1000 events.

\begin{figure}[t]
\includegraphics[height=2.1in]{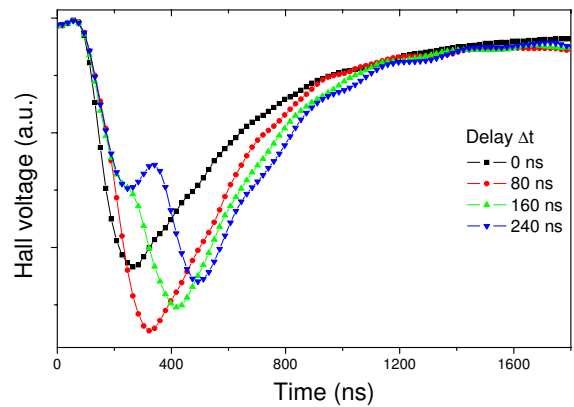}
\caption{\label{fig3} (Color online)
Typical measurement of the Hall voltage as a function of time in an experiment at T=2~K with two microwave pulses with pulse lengths 15~ns separated by a delay $\Delta t$ of a few nanoseconds (the first pulse always starts at t=0). The magnetization of the sample decreases due to the irradiation with microwaves. After the second microwave pulse the magnetization atteins a minimum and relaxes back to the equilibrium value $V_{eq}$.
Note that only every fifteenth datapoint is shown.
}
\end{figure}


Fig. \ref{fig2} shows the magnetization decrease $\Delta M$ during microwave irradiation as a function of the applied magnetic field.
This plot allows us to identify the different resonances according to the microwave frequency.
The first resonance for $f=118$~GHz ($\rm m_s=-10$ to $\rm m_s=-9$) and also $f=107$~GHz ($\rm m_s=-9$ to $\rm m_s=-8$) is located at a magnetic field $\rm \mu_0 H = 0.2$~T.
The detection of transitions between spin states with pulsed microwaves in terms of measuring $\Delta M$ is possible even for very short microwave pulses of the order of several nanoseconds.
In the experiments described below we will always excite the spin system at this first resonance at a magnetic field $\rm \mu_0 H = 0.2$~T.

Fig. \ref{fig3} shows a set of time-resolved magnetization measurements from a typical pump-probe experiment.
The delay of the two microwave pulses is varied in the range of zero to several hundereds of nanoseconds. 
The corresponding time-resolved plots show a pronounced minimum in the magnetization due to the second microwave pulse.
Several features, such as the quite smooth changes of magnetization at the beginning and at the end of the microwave pulses or the rather slow relaxation of magnetization after the two pulses are likely due to a limited response time of the Hall sensor. 

\begin{figure}[t]
\includegraphics[height=2.2in]{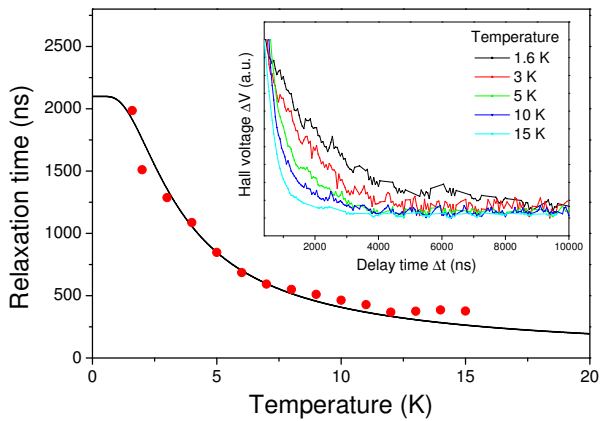}
\caption{\label{fig4}(Color online)
Pump-probe measurement with two microwave pulses with frequencies $f_1 = 118$~GHz and $f_2 = 107$~GHz at the first transition at $\rm \mu_0 H=0.2$~T.
The second pulse (probe-pulse) has always a length of 20~ns, whereas the length of the first pulse (pump-pulse) has a length of 50 ns
In the inset the maximum Hall voltage decrease $\Delta V$ is plotted as a function of the delay between the two microwave pulses for different temperatures.
We observed an exponential decay of $\Delta V$ as a function of the delay time $\Delta t$.
The temperature dependence of the relaxation time, i.e. the energy level lifetime of $m_s=-9$ is in good agreement with theoretical calculations (solid line).
}
\end{figure}

In plotting the response to the second microwave pulse $\Delta V$ as a function of the delay time $\Delta t$, as shown in Fig. \ref{fig4}, we obtain information about the spin dynamics on a nanosecond scale.
The first microwave pulse $\rm p_1$ excites spins from an energy level $\rm m_s=-10$ to the excited energy level $\rm m_s=-9$.
The second microwave pulse $\rm p_2$ excites the spins from the excited level $\rm m_s=-9$ to an even higher level $\rm m_s=-8$.
During the delay time between the pulses the spins can relax back to the ground state on a characteristic time scale, i.e. the energy level lifetime $\tau_m$.
As the response to $\rm p_2$ is directly proportional to the amount of spins in the excited level $\rm m_s=-9$, we can observe the relaxation of the spins from $\rm m_s=-9$ to $\rm m_s=-10$ when plotting $\Delta V$ as a function of the delay time $\Delta t$.
When fitting the experimental data we observed an exponential decay of $\Delta V$ as a function of $\Delta t$ with a caracteristic decay time between 2~$\rm \mu$s and 400~ns with a strong temperature dependence.

\begin{figure}[t]
\includegraphics[height=2.2in]{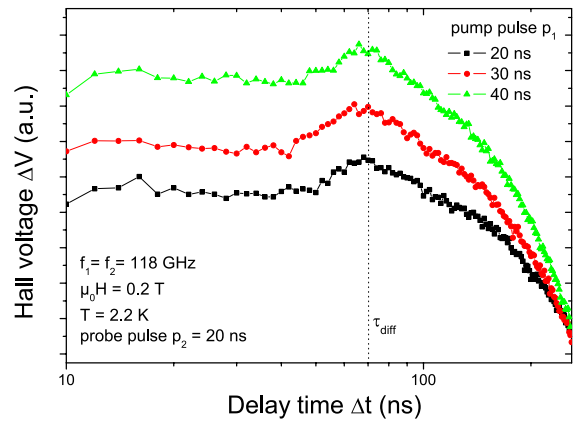}
\caption{\label{fig5} (Color online)
\emph{Hole burning} experiment with one microwave frequency $f=118$~GHz measured at a magnetic field $\mu_0 H=0.2$~T and temperature $T=2.2$~K.
We plotted $\Delta V$ as a function of the delay time $\Delta t$ between the pulses for different pump pulse lengths and probe pulse length 20~ns. 
We observe a characteristic maximum on all the three plots at approximately $\Delta t \approx 70$~ns.}
\end{figure}

The obtained energy level lifetime $\tau_{-9}$ at different temperatures is in good agreement with already published experimental and theoretical results.
\cite {loss:prb2000,furukawa:prb2001}
In the work of \emph{Leuenberger et al.} and \emph{Furukawa et al.} an approach with a general master equation that treats phonon-induced spin transitions between nearest and next-nearest energy levels leads to an equation of the energy level lifetime as a function of temperature and magnetic field, where the only free parameter is the sound velocity in the crystal. 
A fit of our experimental data (black line in figure 4) leads to a sound velocity $v=1580 ~\frac{m}{s}$, which is in good agreement to the expected value in $\rm Fe_8$. 
\cite{furukawa:prb2001}

The above presented pump-probe technique is only sensitive to a small part of the spins that fullfill the resonant condition.
Due to spin-spin interactions the energy absorbed by the molecules in resonance diffuses to molecules that are not in resonance.
This energy diffusion inside the crystal can be studied by a pump-probe technique employing only one single microwave frequency.\cite{wernsdorfer:prb2005}
The first pulse with frequency $\rm f=118$~GHz excites a fraction of spins that are in resonance from the ground state $\rm m_s=-10$ to $\rm m_s=-9$ at $\rm \mu_0 H=0.2$~T.
The second pulse, after a certain time delay $\Delta t$, probes the evolution of the excited spins.
In case of an inhomogneous broadened absorption line, the second pulse probes whether the burnt hole in the spin distribution of the first pulse evolved during the delay time $\Delta t$.
Fig. \ref{fig5} shows the response of the second pulse as a function of the delay between the pulses.
It can be clearly seen, that $\Delta V$ is increasing in the range from $\Delta t$=0 to 70~ns, suggesting that the energy diffusion time inside the crystal is around 70~ns.
In our experiments no temperature dependence of the energy diffusion time was observed in the temperature interval 1.6 to 3~K.
For longer delay times $\Delta t>70$~ns the pulses start to get independent and therefore $\Delta V$ starts to decrease.

The observed energy diffusion time that acts on a very short time scale is mainly due to intermolecular, dipole and hyperfine interactions.
The dipolar and hyperfine coupling in the $\rm Fe_8$ molecule in a crystalline sample has been investigated in numerous publications \cite{wernsdorfer:1999,wernsdorfer:2000Fe8N,tupitsyn:2004,morello:2006}.
When converting the observed time scale $\tau_{diff}=70$~ns to a caracteristic energy ($E=h/\tau_{diff}$) and assuming this energy is the dipolar interaction energy of a macroscopic spin $s=10$ in a magnetic field, we obtain a field $B\approx 1$~mT.
This value suggests that the energy diffusion is dominated by relaxation processes to nuclear spins rather than intermolecular dipolar intercation.
Indeed, when energy resonantly diffuses from spin to spin, a small shift in the energy can occur.
In general, this shift can be provided either by spin-phonon, dipole or hyperfine interactions.
However, the number of low-energy phonons is very small in SMMs.
Energy diffusion due to dipole interactions involve many spins simultaneously.
It has thus a small probability.
Therefore, the hyperfine interaction seems to be the most favourable relaxation process because it involves mainly three bodies (two spins and one nuclear spin).

In conclusion we presented \emph{pump-probe} and \emph{hole burning} experiments on the SMM $\rm Fe_8$ with pulsed microwaves at 107 and 118~GHz.
These experiments allowed us to investigate the level lifetimes of excited spin states and energy diffusion effects, that are both found to be important in terms of spin relaxation.
The level lifetime of the first excited state $\tau_{-9}$ was found to be about 1~$\rm \mu$s, with a strong temperature dependence that can be well explained with general spin-phonon coupling theory.
The energy diffussion time in the crystal was studied with \emph{hole burning} experiments and was determined to be about 70~ns.

\begin{acknowledgments}
The samples for the investigations were kindly provided by A.~Cornia. 
This work is partially financed by EC-RTN-QUEMOLNA Contract No.
MRTN-CT-2003-504880 and MAGMANet. 
We thank E. Eyraud, J.-S. Pelle, J. Florentin and A.-L. Barra for technical and scientific support.
\end{acknowledgments}

\end{document}